\documentstyle[aps,pre,graphicx,multicol,amsmath,amsfonts]{revtex}

\newcommand{\bmulticol}{\begin{multicols}{2}\narrowtext}
\newcommand{\emulticol}{\end{multicols}\widetext}

\begin{document}

\author{Juan M.R.~Parrondo$^{\text{1}}$, Gregory P.~Harmer$^{\text{2}}$,
Derek Abbott$^{\text{2}}$}

\address{$^{\text{1}}$Dep F\'{\i}sica At\'{o}mica,   Molecular y Nuclear, Universidad
Complutense de Madrid, 28040-Madrid, Spain. \\$^{\text{2}}$Centre for Biomedical
Engineering (CBME) and Department of Electrical \& Electronic Engineering,\\
University of Adelaide, SA 5005, Australia.}

\title{New paradoxical games based on Brownian ratchets}

\maketitle

\begin{abstract}
Based on Brownian ratchets, a counter-intuitive phenomenon has recently emerged
-- namely, that two losing games can yield, when combined, a paradoxical
tendency to win. A restriction of this phenomenon is that the rules depend on
the current capital of the player. Here we present new games
where all the rules depend only on the history of the game and not on the
capital. This new history-dependent structure significantly increases the
parameter space for which the effect operates.
\end{abstract}

\bmulticol

In the early 1990's it was shown  that a Brownian particle in a
periodic and asymmetric potential moves to the right (say) in a
systematic way when the potential is switched on and off, either
periodically or randomly~\cite{ajdari93,astumian94}. This so-called
{\it flashing ratchet} is in the class of phenomena known as
Brownian ratchets~\cite{doering95}.
The flashing ratchet can be viewed as the combination of two
dynamics: Brownian motion in an asymmetric potential and
Brownian motion on a flat potential. In each of these two
cases, the particle does not exhibit any systematic motion.
However, when they are alternated the particle moves to the right. The
effect persists even if we add a uniform external force pointing
to the left. In that case, the  two dynamics discussed above yield
motion to the left but when they are combined, the particle moves
to the right.

It has recently been shown, in the seminal
papers~\cite{harmer99a,harmer99b,harmer00a,broeck99}, that a discrete-time
version of the flashing ratchet can be interpreted as simple gambling
games.  Here we have two losing games which become winning when
combined. These games are the simplest situation of a paradoxical
mechanism which, we believe, can be present in many situations of
interest. The apparent paradox points out that if one combines two dynamics
in which a given variable decreases the same variable can
increase in the resulting dynamics. Examples of related phenomena include
enzyme transport analyzed by a four-state rate model~\cite{westerhoff86},
finance models where capital grows by investing in an asset with {\it negative}
typical growth rate~\cite{maslov98}, stability produced by combining
unstable systems~\cite{allison2000}, counter-intuitive drift in the physics
of granular flow~\cite{rosato87}, the combination of declining branching
processes producing an increase~\cite{key87} and counter-intuitive drift
in switched diffusion processes in random media~\cite{pinsky92}.

The games originally described
in~\cite{harmer99a,harmer99b,harmer00a,broeck99} are expressed in terms of
tossing biased coins. The games rely on a state-dependent rule based on the
player's  capital and two losing games can surprisingly combine to win. This
effect was shown to be
essentially a discrete-time Brownian ratchet~\cite{harmer99a}. This is of
interest to information theorists who have long studied the problem of
producing a fair game from biased coins~\cite{gargamo99} and winning
games from fair games~\cite{durrett91}, inspired by the work of von
Neumann~\cite{neumann51} -- the games we are discussing go a step further,
demonstrating a winning expectation  produced from losing games and have
recently been analyzed from the point of view of information theory~\cite{pearce00}.
 Seigman~\cite{seig,key00}
has reinterpreted the capital of the games in terms of electron occupancies
in energy levels, recasting the problem in terms of rate equations.
Similarly, Van den Broeck {\it et al}~\cite{broeck99} have likened the analysis
of the transition probabilities of the games to Onsager's treatment of reaction
rates in circular chemical reactions~\cite{onsager31}.  It has been
suggested in~\cite{harmer00a} that an area of interest
to quantum information theory would be to recast the games in term of
quantum probability amplitudes along the lines
of~\cite{eisert99,goldenberg99,meyer99}. Quantum ratchets have now been
experimentally realised~\cite{linke99}
and thus quantum game theory based on ratchets is of interest.

However, one of the limitations the game paradox and its applicability to
further situations is that it relies on a modulo rule based on the
capital of the player. The modulo arithmetic rule is quite natural for an
interpretation of the paradox in terms of energy levels, say, however for
processes in biology and biophysics it is unnatural. Applicability of the
paradox to population genetics, evolution and economics has been
suggested~\cite{mcclintock99} and thus a desirable version of the paradox would
be to have rules independent of capital.

In this letter we present a new interpretation of the paradox in
terms of good and bad biased coins which are played more or less often
when the two games are combined. This interpretation allows us to
introduce an important modification to the original games, namely,
games which do not depend on the capital but only on the recent
history of wins and loses.

The two original games are as follows. The player has some capital
$X(t)$, $t=0,1,2,\ldots$ In game A the capital is increased by
one with probability $p$ and decreased by one with probability
$1-p$. In game B, the rules are:

\begin{center}
\begin{tabular}{c|c|c}
 & Prob. of win & Prob. of loss \\\hline
 $X(t)/3\in {\mathbb Z}$ &
 $p_1$ & $1-p_1$ \\
$X(t)/3\notin {\mathbb Z}$ &
 $p_2$ & $1-p_2$  \end{tabular}
\end{center}

Here ``win'' means increasing the capital by one and ``loss''
decreasing it by one. For the choice, $p=1/2-\epsilon$,
$p_1=1/10-\epsilon$, and $p_2=3/4-\epsilon$, with $\epsilon >0$,
the two games have a tendency to lose. More precisely $\langle
X(t)\rangle$ is a decreasing function of the number of runs $t$.
However, if in each run we randomly choose the game we play, then,
for $\epsilon$ small enough, $\langle X(t)\rangle$ is an
increasing function of $t$.


An explanation of this paradox is as follows. First, let us
imagine the above rules as implemented by three biased
coins, $A$, $B_1$ and $B_2$, with probability for tails
$p$, $p_1$ and $p_2$, respectively. We see that $A$ and
$B_1$ are ``bad coins,'' whereas $B_2$ is a ``good coin''
for the player. When game B is played alone, at first sight
one would say that $B_1$ is used one third of the time.
However, this is not the case. When the capital is multiple
of three, $X(t)=3n$, there is a high probability of losing,
i.e., $X(t+1)=3n-1$ is the most likely value for the
capital at $t+1$. If this is the case, we have to use coin
$B_2$ in the $t+1$ run and the most likely outcome is now a
win. Therefore, the most likely capital at $t+2$ is again
$X(t+2)=3n$. We see that the probability of $X(t)$ being
multiple of three is bigger than $1/3$, due to the very
rules of game B. The precise value of the equilibrium
probability can be calculated by defining the Markov
process $Y(t)\equiv X(t) \mod 3$, which only takes on three
values, $Y(t)=0,1,2$. The stationary distribution for
$Y(t)$, when $\epsilon=0$ is given by:
$\pi_0=\frac{5}{13};\quad \pi_1=\frac{2}{13};\quad
\pi_2=\frac{6}{13}$. The fairness of the game is indicated
by $\pi_0p_1+ (\pi_1+\pi_2)p_2=1/2$.

When coin $A$ comes to play, the stationary distribution
changes. For instance, if games A and B are switched at
random, one has: $\pi'_0= \frac{245}{709};\quad
\pi'_1=\frac{180}{709};\quad \pi'_2=\frac{284}{709}$. The
game is no longer fair because $\pi'_0=245/709=0.346$ is
closer to $1/3$ than $\pi_0=0.385$, for the ``bad coin''
and now the ``good coin,'' $B_2$, is played more often than
before. The effect persists even if coin $A$ is bad,
leading to the paradox.

This interpretation helps us to find a new version of the paradox
with capital independent games. Game A is the same as before and
we introduce game B$'$ which is played with four coins: $B_1'$,
$B_2'$, $B'_3$, and $B'_4$. Which coin is used now depends on the
history of the game:

\begin{center}
\begin{tabular}{c|c|c|c|c}
 Before last & Last& Coin & Prob. of win & Prob. of loss \\
$t-2$        & $t-1$& ~ & at $t$ & at $t$ \\\hline
loss & loss & $B_1'$ & $p_1$ & $1-p_1$\\ loss & win & $B_2'$ & $p_2$
& $1-p_2$ \\ win & loss & $B_3'$ & $p_3$ & $1-p_3$ \\ win & win &
$B_4'$ & $p_4$ & $1-p_4$ \\
\end{tabular}
\end{center}

This is in fact the most general game depending on the outcome of
the  two last runs. The paradox could even be reproduced with this type
of game if  the ``bad'' coins in game B$'$ are played more often
than what is expected in a completely random game, ie.~one
quarter of the time.

Notice that the capital $X(t)$ in game B$'$ is not a
Markovian process. However, one can define the vector
\begin{equation}
Y(t)=\left( \begin{array}{c} X(t)-X(t-1) \\ X(t-1) - X(t-2)
\end{array}\right)\label{ydef}
\end{equation}
which can take four values $(\pm 1,\pm 1)$, and does form a
Markov chain. The transition probabilities are easily
obtained from the rules of game B$'$.
Let $\pi_1(t)$, $\pi_2(t)$, $\pi_3(t)$ and $\pi_4(t)$ be
the probabilities that $Y(t)$ is $(-1,-1)$, $(1,-1)$,
$(-1,1)$, and $(1,1)$, respectively. The probability
distribution ${\vec \pi}(t)$ verifies the evolution
equation: ${\vec \pi}(t+1)= {\bf A}{\vec \pi}(t)$, where
the matrix ${\bf A}$ is given by the transition
probabilities and reads:
\begin{equation} {\bf A}=\left(
\begin{array}{cccc}
1-p_1 & 0 & 1-p_3 & 0 \\ p_1 & 0 & p_3 & 0 \\ 0 & 1-p_2 & 0 &
1-p_4
\\ 0 & p_2 & 0 & p_4\end{array} \right).
\end{equation}

The stationary distribution ${\vec \pi}_{\rm st}$ of this
Markov chain is by definition invariant under the action of
the matrix ${\bf A}$, i.e., ${\vec \pi}_{\rm st}{\bf
A}={\vec \pi}_{\rm st}$. This distribution reads:
\begin{equation}
{\vec \pi}_{\rm st}=\frac{1}{N}\left(
\begin{array}{c}
(1-p_3)(1-p_4) \\ (1-p_4)p_1 \\ (1-p_4)p_1 \\ p_1p_2
\end{array}\right)
\end{equation}
where $N$ is a normalization constant.

In the stationary regime, the probability to win in a
generic run is:
\begin{equation} p_{\rm win}=
\sum_{i=1}^4\pi_{{\rm st},
i}p_i=\frac{p_1(p_2+1-p_4)}{(1-p_4)(2p_1+1-p_3)+p_1p_2}
\label{pwin}
\end{equation}
which can be rewritten as $p_{\rm win} =1/(2+c/s)$, with
$s=p_1(p_2+1-p_4)>0$ for any choice of the rules, and
$c=(1-p_4)(1-p_3)-p_1p_2.$

Therefore, the tendency of game B$'$ obeys the following
rule: if $c < 0$,  B$'$ is winning; if $c= 0$, B$'$ is
fair; and if $c> 0$, B$'$ is losing. Again, here losing,
winning and fair means that $\langle X(t)\rangle$ is,
respectively, a decreasing, increasing or constant function
of $t$.

Since when game B$'$ is combined with game A the vector
$Y(t)$ as defined in Eq.~(\ref{ydef}) is still a Markov
chain, the same procedure applies. The probabilities of
winning are now replaced by $p_i'=(p_i+p)/2$.
Summarizing, to reproduce the paradox with capital
independent games we have to find a set of five numbers,
$p$ and $p_i$ ($i=1,2,3,4$), such that
\begin{eqnarray}
1-p & > & p\nonumber   \\ (1-p_4)(1-p_3)& > & p_1p_2
\nonumber  \\ (2 -p_4-p)(2-p_3-p) & < & (p_1+p)(p_2+p),
\label{ineqs}
\end{eqnarray}
where the third equation is just the second with $p_i'$ and
the inequality reversed (to make the combined game winning
instead of losing).

One of the coins in game B$'$ must be ``bad'' and used more
often than one quarter of the time. It cannot be either
$B_1'$ or $B_4'$ because the probability of using these
coins depends on whether the game is losing or winning (if
$B_1'$ is played more often than $B_4'$, it is obvious that
the game is losing). The bad coins should be $B_2'$ and
$B_3'$. Let us set $p=1/2-\epsilon$, $p_1=9/10-\epsilon$,
$p_2=p_3=1/4-\epsilon$, and $p_4=7/10-\epsilon$. With these
numbers, one can see that the two first  inequalities in
Eq.~(\ref{ineqs}) are always satisfied if $\epsilon>0$,
whereas the third is satisfied if $\epsilon<1/168=0.00595$
-- ie.~the paradox occurs when $0 < \epsilon < 1/168$, for
our chosen parameter set in this example.

\begin{figure}[htbp]
\begin{center}
\resizebox{6.5cm}{!}{\includegraphics{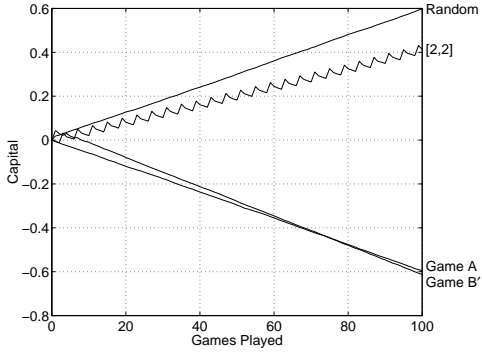}}
\end{center}
\caption{Evolution of capital with play. The lower two
curves show that games A and B$'$ lose when individually
played. [2,2] indicates game A played 2 times followed by
game B$'$ played 2 times and so on.  The top curve
indicates random switching between games A and B$'$.
Capital surprisingly increases in the random or periodic
cases. Simulations are carried out with $\epsilon=0.003$,
with averaging over 500 000 ensembles. $p=1/2-\epsilon$,
$p_1=9/10-\epsilon$, $p_2=p_3=1/4-\epsilon$, and
$p_4=7/10-\epsilon$. There are four possible initial
conditions ---these affect the offsets but not the
slopes--- all the above curves are the average of the four
cases} \label{evolution}
\end{figure}

The simulation in Fig.~\ref{evolution} shows that as games
A and B$'$ evolve individually the capital declines, as
expected (ie.~they are losing games). On the same graph we
see the remarkable result that when A and B$'$ are
alternated either randomly or periodically, the capital now
increases. This reproduces the paradoxical behavior first
observed in the original games~\cite{harmer99a}, but now
without state dependence on capital. The slopes of the
curves corresponding to game B$'$ and to the random
combination can easily be calculated as:
$\langle{X(t+1)}\rangle - \langle{X(t)}\rangle = 2p_{\rm
win}-1$, with $p_{\rm win}$ given by Eq.~(\ref{pwin}).
The old and new games have a fundamental difference in that
the old ones can be interpreted in terms of a random walk
in a periodic environment (RWPE)~\cite{key00} or a Brownian
particle in a periodic potential, whereas the rules of the
present games are homogeneous. We could say that the
periodic structure of the original games has been
transferred to the memory of the rules in the new games.
Therefore, the paradox needs at least one of these two
ingredients: inhomogeneity or non-markovianity\cite{note}.

\begin{figure}[htbp]
\begin{center}
\resizebox{7.5cm}{!}{\includegraphics{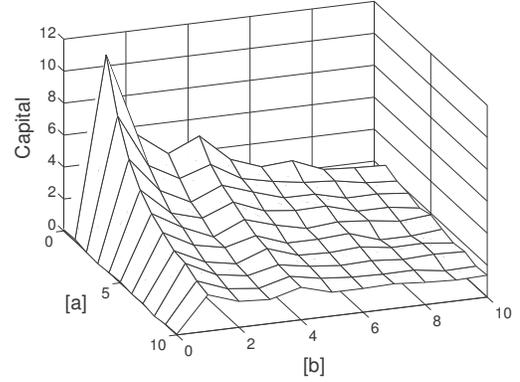}}
\end{center}
\caption{Value of capital after 500 games. Games A and B$'$
are periodically mixed. Game A is played $a$ times,
followed by B$'$ played $b$ times and so on. Games are
played with $\epsilon$=0 and 500 000 ensemble averages have
been taken. $p=1/2$, $p_1=9/10$, $p_2=p_3=1/4$, and
$p_4=7/10$. } \label{periodic}
\end{figure}

Consider now a periodic combination of games A and B$'$. Fig.~\ref{periodic}
shows the capital after 500 games -- where game A is played $a$ times and game
B$'$ is played $b$ times. We can observe that the resulting capital is
greater when the games are switched more frequently. This behavior agrees with
with that of the original games~\cite{harmer99a}. Note that in Fig.~\ref{periodic}
changing the value of $\epsilon$ only affects the vertical capital displacement,
thus setting $\epsilon=0$ pushes the graph into the positive region.

\begin{figure}[htbp]
\begin{center}
\resizebox{8cm}{!}{\includegraphics{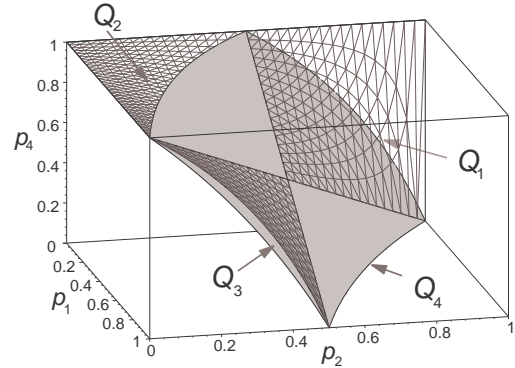}}
\end{center}
\caption{Parameter space for game B$'$ when $p$=1/2. We see
there are four volumes labelled $Q_1$, $Q_2$, $Q_3$ and
$Q_4$, bounded by the inequalities in Eq.~(\ref{ineqs}).
The paradox of two losing games that win if randomly
combined, occurs if the parameters lie within the volumes
marked $Q_1$ and $Q_3$. In regions $Q_2$ and $Q_4$ the
reverse effect occurs where games A and B are individually
winning, but the randomized combination is losing. }
\label{space}
\end{figure}

For the randomized games, we can now observe the volume of
parameter space for which the paradox takes effect, by
plotting the surfaces that represent the boundaries of the
inequalities in Eq.~(\ref{ineqs}). This is shown in
Fig.~\ref{space}, where for convenience we have set
$p_2=p_3$ to produce the graph in three variables. The
volumes enclosed by the surfaces marked $Q_1,{\ldots},Q_4$
are the regions of parameter space for which the paradox
takes effect. Regions $Q_1$ and $Q_3$ are where two losing
games combine to win. On the other hand, $Q_2$ and $Q_4$
represent the reverse effect where two winning games
combine to lose.
This conjugate region can be simply thought of in terms of
changing the sign of the capital,  so that the perspective
of the concepts `win' and `lose' reverse.
This was observed
in the original capital-dependent games~\cite{harmer00b},
however the conjugate regions were symmetrical. What is now
interesting is that the new history-dependent games have
asymmetrical conjugate regions, as can be seen in
Fig.~\ref{space}.


Another important comparison between the new
history-dependent games and the original capital-dependent
games is that the volume of parameter space is now bigger.
A numerical mesh analysis on Fig.~\ref{space} revealed that
the new games have a parameter space about 50 times larger
than the original games reported in~\cite{harmer00b}. For
applications such as in biophysics, it is important to find
such gaming models with large and hence robust parameter
spaces. Although it appears that the rates of winning from
the slopes of Fig.~\ref{evolution} are about factor of 2
lower than the original games, this is only the case for
the particular chosen parameters. The 50 times increase in
parameter space is favorable for applications in modeling
evolutionary processes in biology, for example, where a
weak pay-off can gradually accumulate over a long period of
time.

In summary, we have shown that the apparently paradoxical
effect where two losing games can cooperate to win does
work with a history based state-dependent rule rather than
the original restriction of a modulo capital based
state-dependence. This, together with an increased
parameter space opens up the phenomenon to a wider range of
possible application areas. This suggests that future
investigation of further types of history-based rules and
other types of state dependencies may be fruitful.

This work was supported by the Direcci\'on General de
Ense\~nanza Superior e Investigaci\'on Cient\'{\i}fica
(Spain) Project No.~PB97-0076-C02, GTECH (USA), The Sir
Ross and Sir Keith Smith Fund (Australia) and the
Australian Research Council (ARC).


\emulticol


\begin{thebibliography}{99}
\bibitem{note} One could design a simpler non
Markovian game depending only on $X(t)-X(t-1)$, i.e., the
outcome of the last run. However, this memory is too short
to induce any paradoxical behavior. The reason is that in a
fair game with such one-step memory the rules are always
invariant under the change $X(t)\to -X(t)$, and this
symmetry must be broken to have the paradox. A similar
situation occured in the original games where a modulo 3
rule was necessary since a modulo 2 rule is not enough to
break the symmetry $X(t)\to -X(t)$.
\end{thebibliography}
\end{document}